\documentclass{article}

\usepackage{arxiv}

\usepackage[utf8]{inputenc} 
\usepackage[T1]{fontenc}    
\usepackage{hyperref}       
\usepackage{url}            
\usepackage{booktabs}       
\usepackage{amsfonts}       
\usepackage{nicefrac}       
\usepackage{microtype}      
\usepackage{lipsum}		
\usepackage{graphicx}
\usepackage{natbib}
\usepackage{doi}
\usepackage{comment}

\title{Towards Universal Criteria for Machine Consciousness: AI, Shared Essence, and Neuroscience}

\author{ \href{https://orcid.org/0000-0002-7612-9244}{\includegraphics[scale=0.06]{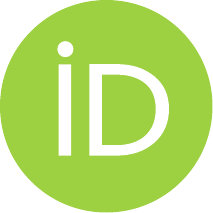}\hspace{1mm}Nur Aizaan Anwar }
  \\
	Department of Computing and Department of Surgery \& Cancer\\
	Imperial College London\\
	London SW7 2AZ\\
	\texttt{aa10920@ic.ac.uk} \\
	\And
	\href{https://orcid.org/0000-0000-0000-0000}{\includegraphics[scale=0.06]{orcid.pdf}\hspace{1mm}Cosmin Badea} \\
	Department of Computing\\
	Imperial College London\\
	London SW7 2AZ\\
	\texttt{c.badea@imperial.ac.uk} \\
}

\renewcommand{\headeright}{}
\renewcommand{\undertitle}{}
\renewcommand{\shorttitle}{Towards Universal Criteria for Machine Consciousness}

\hypersetup{
pdftitle={Universal Criteria for Machine Consciousness},
pdfauthor={Nur Aizaan Anwar, Cosmin Badea},
pdfkeywords={Consciousness, Machine Consciousness, Artificial Intelligence, Machine Intelligence, Neuroscience of Consciousness, Philosophy of Artificial Intelligence},
}

\begin{document}
\maketitle

\begin{abstract}
As artificially intelligent systems become more anthropomorphic and pervasive, and their potential impact on humanity more urgent, discussions about the possibility of machine consciousness have significantly intensified, and it is sometimes seen as `the holy grail'. Many concerns have been voiced about the ramifications of creating an artificial conscious entity. This is compounded by a marked lack of consensus around what constitutes consciousness and by an absence of a universal set of criteria for determining consciousness. By going into depth on the foundations and characteristics of consciousness, we propose five criteria for determining whether a machine is conscious, which can also be applied more generally to any entity. This paper aims to serve as a primer and stepping stone for researchers of consciousness, be they in philosophy, computer science, medicine, or any other field, to further pursue this holy grail of philosophy, neuroscience and artificial intelligence.

\end{abstract}


\paragraph{Acknowledgement of Funding}
This work was supported by the UKRI CDT in AI for Healthcare http://ai4health.io (Grant No. EP/S023283/1)

\newpage

\section{Introduction}
Of late, there has been a significant surge of interest and importance awarded to the issue of machine consciousness, especially given the explosive recent developments in artificial intelligence \citep{2023arXiv231003715M}, including the development of decision-making artificial agents \citep{Badea2021HaveAB}, embodied or not, and the rise of anthropomorphic automated systems \citep{Tiku2022GoogleLife, Urwin2023TopNow, AguerayArcas2022ArtificialArcas, Chalmers2023CouldConscious}. We shall from now on use the term `machine' to mean all such entities, irrespective of the particularities of their physical composition.

In the current landscape of explosive development in artificial intelligence, what has reignited the debate around consciousness is the seemingly imminent possibility of creating and deploying highly developed, potentially conscious, machines. The potential moral and legal implications of this, and the consequences or potential harm these entities may inflict upon humanity, whether individually \cite{Hindocha2022MoralHealthcare}  or collectively \citep{Agar2020HowMinds, Hildt2023TheConcerns}, have made for an unprecedented rise in academic, industrial, political, as well as public, interest in machine intelligence and consciousness \citep{Federspiel2023ThreatsExistence}. Therefore, as the usage of artificially intelligent systems becomes more and more pervasive, it is clear that the need to understand their capacity and potential for consciousness, as well as to develop criteria for deciding upon such a capacity, become crucial. 

Questions about this are of interest to many different fields, both the more theoretical and the more applied, including computer science, cognitive science, philosophy of mind and medicine. Some of the most important such questions about the behaviour of machines which we have examined in a similar vein before include: \begin{itemize}
    \item how we ought to build them \citep{Seeamber2023IfSo, Badea2022EstablishingMF}
   \item how we ought to ensure they obey the rules we give them \citep{Badea2021MoralityMA} and what their capacity for moral behaviour is \citep{Badea2021HaveAB}
   \item how we could interpret the nature and outcomes of their behaviour \citep{Vijayaraghavan2023MinimumLO}
   \item what their role and relation to us, actual or desired, might be \citep{Hindocha2022MoralHealthcare}
   \item what the impact they could or should have in the domain in which they are deployed is \citep{Bolton2022DevelopingUse}
\end{itemize}

Unique amongst such questions is \textit{the question of whether machines can be conscious}, a question whose resolution (or lack thereof) would influence most, if not all, of the above topics. Making progress on this question might present us with myriad different benefits and contributions in directions such as: clarifying or dispelling existing assumptions about the nature of such entities, coming up with efficacious and more accurate ways of going about in our consideration of them, and even perhaps identifying promising new theories of consciousness and thus better understanding the mind. This paper aims to make a contribution towards this goal by presenting a set of criteria that can be used to make progress on and hopefully be able to answer this question.

\subsection{What is Consciousness?}

Various perspectives on consciousness exist, yet none of these competing theories has come close to unifying them or achieving consensus \citep{Butlin2023ConsciousnessConsciousness, Seth2022TheoriesConsciousness, Consortium2023AnConsciousness}. What makes it even harder to achieve such a consensus is that at its core lies the issue of how one defines consciousness, and some scholars even purposefully omit to provide a clear definition of it, instead focusing on its properties and the process of its emergence \citep{dennett1991ConsciousnessExplained}. It also warrants mentioning that one could either come up with a theory of consciousness first, and then build a definition for it, or start with a particular definition and construct the theory around it, and this is especially relevant if one considers consciousness to be a form of computation, because then the criteria for determining if something is conscious or not changes significantly \citep{Butlin2023ConsciousnessConsciousness, Tononi2015Consciousness:Everywhere, Krauss2020WillMachines, Searle1980MindsPrograms}. Even in medicine, where the determination of consciousness informs life or death decisions \citep{Post2022BreakingCalculus.}, there exists no definitive definition nor irreproachable test for it \citep{Edlow2023MeasuringUnit, Monti2010WillfulConsciousness, Huang2023FunctionalConsciousness}, highlighting once again the importance and far reach of such work. 

Whilst a definitive definition of consciousness has so far eluded us, we may yet be able to describe ‘how’ consciousness is, with some degree of agreement in the literature, especially orbiting around the subjective, qualitative experience of being: \textit{qualia} \citep{Chalmers1996TheMind, Nagel1974WhatBat}. Neuroscientist Giulio Tononi \citep{Tononi2008ConsciousnessManifesto} and philosopher John Searle \citep{Searle2013OurConsciousness} describe consciousness as “what vanishes when we fall into dreamless sleep and reappears when we wake up or dream”. Physicist Max Tegmark likens consciousness to the way the subject feels internally when information is being processed, suggesting consciousness may be a new state of matter, the `perceptronium'\citep{Tegmark2015ConsciousnessMatter}. We believe that such approaches can help to cut straight to the heart of the matter for an issue that relies on something deeply subjective (qualia) which we would, being scientific in our approach, wish to treat as objectively as possible.

Due to the wealth of different definitions of consciousness, and there being as of yet unresolved conflicts between them, it would be ideal if, in our treatment of the topic of consciousness, we could be flexible enough so that our results apply across this spectrum of definitions. We thus aim, in our criteria below, to cater to any of the above definitions, by remaining agnostic as to which the 'right' one is. We can this because we also, as the work mentioned in the previous paragraph does, focus on examining 'how' consciousness is, and our criteria flow well from considering this angle.

\section{Determining consciousness and the shared essence argument}

\subsection{Consciousness in the Inanimate: The Challenge}

\paragraph{The Shared Essence Argument}
\textit{Prima facie}, we have what appears to be a solid argument from analogy that allows one to assert the existence of consciousness in our fellow man, if we simply rely on direct observation of consciousness in ourselves (as we will see with Descartes' argument discussed below), and add the ingredient of the obvious and extensive similarity in constitution between ourselves and them. In other words, because we are conscious, and they are `like us', they are also conscious. Call this `\textit{the shared essence argument}'. 

However, what of entities that might differ in physical constitution to ourselves, inanimate entities such as machines, for which we cannot immediately use the shared essence argument to claim that they are capable of consciousness? 

Firstly, a `machine' has been described as a physical system that can perform functions \citep{Searle2015ConsciousnessIntelligence}. The term, as we use it here, refers to non-biological systems (not carbon-based). In this paper the term `machine' is not restricted to embodied entities only, as we wish to be agnostic as to the physical composition of the machines we speak of, in order to encompass all artificial agents including abstract ones which have no body, such as large language models \citep{2023arXiv231003715M}. Since their constitution is different, the shared essence argument does not directly apply, so how, then, do we show that a machine is conscious? 

One might think that one can simply modify the shared essence argument to take this into account by, for instance, claiming that machines do have a shared essence to us, and instead of their physical composition it is their functional composition - they do things like we do. However, how could we guarantee that machines are functionally similar to us, that they think or operate, "mentally", in the same way as we do?

Here we have one of the biggest challenges to our endeavour because, strictly speaking, one cannot definitively conclude that any other entity is conscious without having access to their subjective experience (should they have any), so that one may check the presence or absence of consciousness. Since we cannot directly access the subjective experience of another entity, as argued in \citep{Nagel1974WhatBat}, we need another way. What might this way be? Well, due to the difference between the immediate validity of our own claim to consciousness (made to ourselves) and the claim of others (made to us), and seeing as we cannot apply the shared essence argument anymore, \textbf{it must be that others’ consciousness be somehow demonstrable to us}.

This is the strongest rationale for it being essential to devise a set of criteria for use in such a determination or demonstration of others' consciousness, and this is precisely what we are aiming to contribute to with our framework. As we show in this paper, in order to be able to prove that others are conscious, whether they be man or machine, one can indeed verify that several conditions are satisfied, conditions that pertain to the nature of matter and humans’ capability to directly or indirectly detect consciousness in others. 

\subsection{Towards Universal Criteria}

Note that, in line with our arguments above, we cannot in this paper definitively provide the answer to the millennia-old question, reaching as least as far back as Aristotle \citep{Smith2007OnAristotle}, of whether the inanimate can be conscious. However, while exploring historical and contemporary views of consciousness, we will propose a new set of criteria which are assumption-free for determining consciousness in machines. 

Indicators and tests of consciousness have previously been presented in the literature, however these generally presuppose a specific view of what consciousness is in their underlying paradigms, such as in the following two papers which we discuss below \citep{Dung2023TestsConsciousness, Butlin2023ConsciousnessConsciousness}. In the first paper, by extrapolating studies of non-human animals to machines,  \citet{Dung2023TestsConsciousness} proposed that consciousness is a product of sensory input, and can be tested in machines if they are fitted with equipment that can mimic animals' sensory system. However, there is the possibility (called 'gerrymandering') \citep{Tomasik2014DoAR} of artificially obtaining such behaviour not due to the machines' own will, but due to simple conditioning or training, in a way simply imitating the behaviour of the animals, so that it does not actually imply conscious behaviour. Provided that there is no gerrymandering, its (now free) response to stimuli can be observed and compared to a non-human animal, and thus a parallel can be drawn between the two, and therefore an argument be made that if animals can be conscious then so can machines. A limitation to this is the possibility of machine consciousness manifesting not through sensory processes but rather in a completely different way altogether. In another approach, \citet{Butlin2023ConsciousnessConsciousness} proposed a list of 14 indicator properties of machine consciousness, based on the assumption that consciousness arises from the right type of computation, a theory known as `computational functionalism'\citep{Putnam1967PsychophysicalPredicates}, and mental states and events are in fact computational states of the brain. 

In conclusion, most if not all of the frameworks found in the literature are only fully applicable if one commits to a particular paradigm of consciousness, while excluding others, and is therefore not as general as what we wish to achieve here. Thus, we reaffirm our aim to not make any initial assumptions about the nature of consciousness, but rather present sequentially-verifiable conditions (each relying on the validity of the preceding ones) for determining consciousness which can then be applied to any machine or artificial system and which are flexible enough to fit with any conception of consciousness. Therefore our results should be applicable most generally and thus could serve as a basis for investigation in any future work, regardless of the perspective taken on consciousness.

\section{Criteria for determining machine consciousness}


\subsection{Criterion 1: Consciousness must exist}

Before making any claim, note that consciousness must of course, on one way or another, first exist as otherwise, this question of determining the consciousness of a machine cannot even be meaningfully considered. Well, if consciousness (as a subjective experience) is not directly observable, as argued above, how can I, personally, know that it is real? Descartes' work suggests that this can indeed be done through rational means, and through his \textit{"dubito, ergo cogito, ergo sum"} ("I doubt, therefore I think, therefore I am"), together with the reflective and self-referential nature of thinking, I can at least conclude that I myself am conscious\citep{Descartes1637DiscourseSciences}. Philosophers like Daniel Dennett \citep{Dennett2003TheConsciousness} and Keith Frankish \citep{Frankish2019WhatEssays} have attempted to refute this argument, by claiming that consciousness is an illusion created by our physical body. 


By using examples such as optical illusions which are not corrected even after explanation, Dennett and Frankish attempted to show that what we perceive is what our physical body has created as a representation only, and not actual reality. However, the question of whether what we perceive is a false representation or not is another matter ("\textit{the veil of perception}"), as perceive we still do; and, we believe, to think about what we perceive constitutes and thus demonstrates the act of deliberation, and thus of consciousness. Furthermore, upon introspecting upon introspection itself, one can even claim that consciousness is the most unquestionable existence there is \citep{Chalmers2020DebunkingConsciousnessb}, which is something Dennett and Frankish would challenge. 


\subsection{Criterion 2: I am not the only conscious entity to exist}

The second condition to be met is that I am not the only conscious entity, and that other elements in my universe are also capable of the same feat. Otherwise, the question that this paper seeks to answer cannot be addressed. This conundrum of others' being conscious arose from the Cartesian view of consciousness above, whereby the only surety I ultimately have, is my own existence as a thinking thing. There is truly no direct way to know that other conscious entities exist \citep{Chalmers2013WhyMysterious}, as we have also argued above. Even those with whom I share the most similar constitution and behaviour, and thus in spite of the shared essence argument, are not necessarily conscious. For instance, when I have a vivid dream of other people, do they truly exist, or are they merely figments of my own imagination? Who is to say I am not dreaming now, and I am only imagining that you are reading this paper? Mimicry of human conversation is the basis for the Turing Test \citep{Turing1950ComputingHttp://www.jstor.org/sta} which also what led Google scientists to claim that Google's conversational model might exhibit signs of consciousness \citep{Tiku2022GoogleLife, AguerayArcas2022ArtificialArcas}. 

However, recent observations in clinical neuroscience cast doubt on the shared essence argument. It has been shown that some humans with sleep disorders, loosely coined sleepwalking, are capable of performing acts which appear conscious, such as eating, driving and even committing violent acts \citep{Popat2015WhileParasomnia}, but which are arguably not conscious as they do not occur in full cognitive control or even awareness. Thus the appearance of consciousness might not at all be definitive proof of its presence. For all I know, you, my readers, might all be sleepwalking, while I am forever ‘trapped’ in my cage of consciousness. Relying exclusively on the mimicry of human behaviour for assessing this (`behaviourism') also assumes that those exhibiting non-human behaviour are incapable of consciousness, which, based on animal studies, is contentious as there are conscious-like behaviours in non-human species \citep{Carls-Diamante2022WhereOctopus, Dung2023TestsConsciousness, SETH2005119}. Regardless, for practical reasons, we must from here on assume that other entities are also capable of subjective experience, to allow us further examination of the possibility of machines being conscious. 

\subsection{Criterion 3: Matter is sufficient for consciousness to arise  }

Is matter a sufficient or necessary requirement for consciousness or is it completely independent of consciousness altogether? This "hard problem" of consciousness proposed by David Chalmers is the question of how a physical body can give rise to subjective experience at all \citep{Chalmers2011FacingConsciousness}. He sees conscious experience akin to other fundamental aspects of the universe, like mass or space, and considers that physical processes might not be enough to explain consciousness, which might suggest that matter is not a sufficient or necessary requirement for consciousness. Descartes himself argued for \textit{substance dualism}, namely the perspective that matter or physical substance ("res extensa") exists independently of consciousness or thinking substance ("res cogitans") \citep{Descartes1637DiscourseSciences}. Despite being physiologically similar to non-human animals, he claimed that only humans fully possess consciousness and other animals do not, to the same extent at least, as consciousness is not an intrinsic property of matter, but rather, of humans. Leibniz disagreed, and argued that all forms of matter are capable of consciousness \citep{Leibniz1898TheMonadology}. Yet, he refuted the possibility of matter fully accounting for it, as what is infinitely divisible cannot produce indivisible subjective experience (e.g. the taste of coffee is the way it is, and cannot be experienced in its constituent parts at the same time). This is echoed in contemporary views on emergentism and non-reductive physicalism \citep{Nagel1974WhatBat}, which also espouse this idea that while connected, consciousness cannot fully be reduced to physical substance.

 Neuroscientific and medical phenomena have shown us that matter (our brain) is at least necessary for consciousness. For example, individuals who have sustained unilateral cerebral injuries (without any ocular or muscular impediment) have been known to be completely unaware of the side of their bodies and visual fields opposite to the site of brain damage \citep{Vuilleumier2007HemispatialNeglect}. These patients would not even be conscious (anosognosic) that they are neglecting one side of their bodies and some have been known only to groom one side of their faces, leaving the other unkempt. Another phenomenon supporting the necessity of matter for consciousness is cortical blindness, whereby damage to the cerebrum’s occipital cortex renders an individual blind, in the absence of ocular injury \citep{Rapcsak2018CorticalNeuropsychology}. 

Whilst clinical neuroscience suggests indications of the necessity of matter, it has also done the opposite. The human brain consists of approximately 80-100 billion neurons, and 70\% of them reside in the cerebellum (the second largest component of the brain responsible for movement and coordination) \citep{Herculano-Houzel2009TheBrain}. One could conclude that if matter was necessary for consciousness, then the absence of the cerebellum should likely render an individual with impaired consciousness. However, there have been accounts in the medical literature to show that individuals without cerebella are able to lead normal lives \citep{Lemon2010LifeCerebellum, Yu2015APatient}, suggesting minimal, if any, impairment. Perhaps it is a question of where the matter is, not how much there is. Then we might argue that it is the cerebrum (the largest part of the brain responsible for sensory control, voluntary actions, language, memory and planning) that is responsible for consciousness, not the cerebellum. However, some individuals whose cerebrums have been severely effaced by excessive fluid in the brain (hydrocephalus) have also shown the ability to lead normal lives, even when more than 50-75\%, or even more than 90\%, of their brain volume was gone \citep{Feuillet2007BrainWorker, Lewin1980IsNecessary}. In one study, more than half of those with 90\% or more hydrocephalus had IQ's above 100 (the average). This may be due to the brain's ability to remodel in the face of pathologies (neural plasticity) that seemingly defy compatibility with normal functioning \citep{Ferris2019LifeHydrocephalus, Alders2018VolumetricVentriculomegaly}. 

Yet, as the above studies also highlight, severe impairment can also be observed alongside such a reduced brain volume, and since both impairment and non-impairment can be observed with such variations in brain condition, one could then argue that perhaps the brain is not after all the locus of consciousness. However, medicine has also shown that individuals would still retain their consciousness should extra-cranial organs be destroyed or replaced (even with mechanical substitutes) \citep{Burkle2014WhyConfusion.}. Therefore, it is fair to argue that neuroanatomical accounts of human consciousness suggest that the brain is the strongest candidate for the seat of consciousness, yet this only supports the argument for the necessity of the brain in consciousness, not for its sufficiency. 

Having seen arguments for both sides of the necessity debate, we will now discuss sufficiency. It seems that, to remain as universally applicable as possible, our framework has to avoid committing to the potential that the brain is not sufficient for consciousness. This is because, if we were to commit to this and suggest that some other type of non-physical matter is necessary, we would find ourselves at an impasse in deciding upon the consciousness of potentially purely physical entities (humans or machines). Thus, for our purpose of meaningfully distinguishing consciousness in such entities, we shall avoid further debate around (Cartesian) substance dualism, instead discounting it for now, and assuming a position in which physical matter is sufficient, or more accurately, \textbf{physical matter is not insufficient} to give rise to consciousness. 



\subsection{Criterion 4: Machines must be conducive for consciousness}
Assuming that the physical seat of consciousness has been confirmed in the brain, and this can be artificially constructed, would that be sufficient to give rise to consciousness? Is the structure sufficient or must there be conducive underlying mechanisms also? In humans, we seem to require more than normal anatomy to be conscious. For example, most of us possess intact brains, yet, when we have dreamless sleep, our consciousness disappears while the cerebral cortex remains as active as during wakefulness \citep{Tononi2016IntegratedSubstrate}. Physically normal newborn infants do not develop consciousness until much later in life \citep{Lagercrantz2009TheLife}. Then, in what state must a physical system be for consciousness to arise? In humans, current experimental approaches taken to determine the minimal components required for consciousness (neural correlates of consciousness or NCC) involve comparing neural measures (electro-chemical, anatomical, metabolic and biochemical) during conscious and unconscious states \citep{Lepauvre2021TheChallenges, Friedman2023TheAspects, Chalmers2000WhatConsciousness}. Many of these measures have led to greater understanding of the biological elements related to consciousness, however, there has yet to be a consensus on the minimal conditions giving rise to it. 
Even when we have uncovered the minimal components for consciousness in humans, must machines then be composed of biological matter to be conducive for consciousness? Proponents of the computational basis for consciousness take on the view that any system, regardless of its physical nature, can be conscious if it performs the right form of information processing \citep{Butlin2023ConsciousnessConsciousness, Reggia2016WhatConsciousness}. Similar to the search for the NCC, the minimal computational criteria (correlates) for consciousness have also yet to be found. Nonetheless, for the purpose of progressing to the final condition, we must assume that we have discovered the minimal requirements for consciousness and this has been replicated in machines.

\subsection{Criterion 5: Consciousness must be observable}

We now have a machine made conducive for consciousness. How do we know it is conscious? Even in medicine, where there is a crucial need to determine whether a patient is conscious or not, a reliable consciousness metric remains elusive. At the bedside, doctors still make assessments based on patients' behavioural responses \citep{Farisco2016AdvancingCare, Kondziella2023NeurologicalBedside, RoyalCollegeofPhysicians2020ProlongedInjury, Wade2022ProlongedGuidelines.}. Yet, the reliability of such methods and their rates of diagnosis have been widely contested \citep{Lawrence2023CommunicatingOverview, Wang2020TheAssessment, Schnakers2009DiagnosticAssessment, Friedman2023TheAspects, Scolding2021ProlongedGuidelines}. As we have seen in Condition 2, anthropomorphic behaviour does not necessarily result from consciousness. In addition, activities that humans undertake which have been simulated by artificial intelligence such as face recognition, speech recognition, and processing meaning in words do not necessarily arise from conscious processes \citep{Dehaene2017WhatIt}. Non-behavioural assessments of human consciousness, such as electrophysiology and imaging, may assist in detecting consciousness, yet such methods still require refining and have not been adopted routinely in clinical practice \citep{Kondziella2016PreservedMeta-analysis, Wade2022ProlongedGuidelines.} Even if they were, they cannot directly be applied to non-biological systems.

\section{Conclusion}

This paper has presented and discussed five criteria that an observer can consult in order to decide that another entity is conscious. The issues surrounding them have been debated since time immemorial, remain fundamental and require resolution. Whilst our criteria provide a new framework for determining consciousness, our investigation of them shows the limitations of our knowledge of the true nature of consciousness, and the urgency around further advancing our understanding of it. They highlight the need to progress our knowledge in both biological and non-biological entities, and pave our way towards doing so. By using our criteria and overcoming the barriers we have highlighted in terms of what, how, and where consciousness is, we may finally conclude whether such entities can or cannot be conscious.

\setcitestyle{numbers} 
\bibliographystyle{unsrtnat}
\bibliography{references}  






\end{document}